\documentclass[%
 reprint,superscriptaddress,
 amsmath,amssymb, nofootinbib,
 aps,twocolumn,pra
]{revtex4-1}
\usepackage{graphicx}
\usepackage{multirow}
\usepackage{amsfonts}
\usepackage{float}
\usepackage{bbold} 
\usepackage{verbatim} 
\usepackage{ulem} 
\usepackage[width=.8\textwidth]{caption}
\usepackage[margin=0.01cm]{caption}
\usepackage[usenames, dvipsnames]{xcolor}
\usepackage{xifthen}
\usepackage{orcidlink}

\newcommand{\arxiv}[2][]{\ifthenelse{\isempty{#1}}{\href{http://arxiv.org/abs/#2}{{\tt arXiv:\allowbreak{}#2}}} {\href{http://arxiv.org/abs/#2}{{\tt arXiv:\allowbreak{}#2 [#1]}}}}

\begin{document}

\title{A Compact Quantum Random Number Generator Based on Balanced Detection of Shot Noise}

\author{Jaideep Singh\orcidlink{0009-0003-0703-8837}\smallskip}
\affiliation{Quantum Research Centre, Technology Innovation Institute, Abu Dhabi, United Arab Emirates}
\author{Rodrigo Piera\orcidlink{0000-0001-6072-3057}\smallskip}
\affiliation{Quantum Research Centre, Technology Innovation Institute, Abu Dhabi, United Arab Emirates}
\author{Yury Kurochkin\orcidlink{0000-0001-5376-6358}\smallskip}
\affiliation{Quantum Research Centre, Technology Innovation Institute, Abu Dhabi, United Arab Emirates}
\author{James A. Grieve\orcidlink{0000-0002-2800-8317}\smallskip}
\affiliation{Quantum Research Centre, Technology Innovation Institute, Abu Dhabi, United Arab Emirates}

\begin{abstract}
\noindent Random Number Generators are critical components of modern cryptosystems. Quantum Random Number Generators (QRNG) have emerged to provide high-quality randomness for these applications. Here we describe a scheme to extract random numbers using balanced detection of shot noise from an LED in a commercially available off-the-shelf package. The balanced detection minimizes classical noise contributions from the optical field, improving the isolation of the quantum noise. We present a detailed description and analyze the performance of a QRNG that can be easily integrated into existing systems without the requirement of custom components. The design is optimised for manufacturability, cost, and size.
\end{abstract}

\maketitle

\section{Introduction}

\noindent Randomness generation is an essential ingredient in many applications, spanning cryptography~\cite{randomness_crypto}, gaming~\cite{quantum_lottery}, network management, fundamental physics experiments~\cite{CHSH_randomness} and sampling~\cite{monte_carlo_method,random_algorithms} as a non-exhaustive list. In most of these applications, random values must be both unpredictable and uniformly distributed: for example in quantum key distribution qubit values must be chosen in a way that is random from the perspective of a would-be evesdropper~\cite{quantumtech_telecomm}. In post-quantum cryptography, the security strongly depends on the properties of the random number generator, any correlations in the output pose serious security risks to the protocol~\cite{qrng_cryptography,qrng_review}. Randomness of this type can only be obtained by sampling physical processes that are intrinsically unpredictable. Quantum systems have emerged as compelling candidates for this task, as by utilising quantum random processes it is possible to access unpredictability which does not depend upon ignorance of the system or lack of information. Devices which embrace this approach are known as Quantum Random Number Generators (QRNGs). Popular quantum processes for this task include phase diffusion~\cite{carlosqrngquside,yury_qrng_phase_diff,qrng_phase_fluc}, vacuum fluctuations~\cite{christianqrngvacuum,100gbpsvacuumqrng,qrng_vacuum_fluc} and photonic shot noise~\cite{qrngonmobileidq}, with commercial examples later developed~\cite{ID_Quantique_2022_QRNG,Quside_2024}. Of these systems, photonic shot noise (sometimes labelled ``photon-counting'') schemes have emerged as particularly amenable to industrialisation~\cite{ID_Quantique_2022_QRNG,qrng_review}, and in this manuscript, we describe such an effort.

Many QRNGs based on photonic shot noise do not consider the optical field's stability. The optical field usually consists of both shot noise and classical noise from sources such as thermal sources or power supply noise, which can introduce correlations into the system such as switching noise from power supplies or power supply modulation by an adversary. Typically stability is assumed from the design, however, it is hard to decouple this noise from the shot noise since they are usually spectrally the same on the output or require active compensation.

A balanced detection approach is widely used with lasers to reduce the classical noise in the optical field and make low noise measurements at the shot noise limit~\cite{laser_noise_supression,balanced_photodetector}. This approach however has not been explored for LEDs in the context of QRNGs.

LEDs are widely available light sources and easy to manufacture. These devices can be integrated and do not require complex hardware for control. There has been some research on LED-based QRNGs most significantly from ID quantique and recently on Perovskite LEDs~\cite{led_qrng_entropy,perovskite_led_qrng}. The use of spatial beam splitting reduces the complexity of using discrete components like beamsplitters. Hu et. al. demonstrated a QRNG developed using a phototransistor optocoupler, which can be used to drastically minimise the size, cost and complexity of the device~\cite{Optocoupler_qrng}. Optocouplers are widely available off-the-shelf components, this increases the trust for OEM cryptography device manufacturers since they can procure parts from trusted supply chains to ensure infrastructure security.

In this work, we present a QRNG based on the balanced detection of the optical field of an LED in a linear optocoupler device, to reduce classical noise in the system and isolate the shot noise. We demonstrate a Quantum-to-Classical noise ratio (QCNR) of greater than 30\,dB, in a device realised using entirely off-the-shelf components widely available in the electronics supply chain, and without any non-standard manufacturing steps.

\section{Theory}

\noindent Shot noise is the statistical fluctuation of photon numbers in an optical field due to their independent nature. Many light sources emit photons at random times, so the exact number of photons produced during a specific interval cannot be accurately predicted~\cite{sub-shot-noise-imaging}. This effect is widely known in optical metrology, giving rise to the ``shot-noise limit'', which bounds the precision with which one can measure optical intensity~\cite{shot-noise-detector,shot-noise-photodiode-ch-fox}. Only specially prepared light fields such as amplitude-squeezed light can overcome this limit~\cite{squeezed-light}. In our scheme, we consider our light source as trusted. The photon number statistics in this case for shot noise are described by a Poisson distribution. We first describe how our scheme isolates shot noise, before discussing how shot noise can be detected and used for our QRNG.

\begin{figure}
    \centering
    \includegraphics[width=\linewidth]{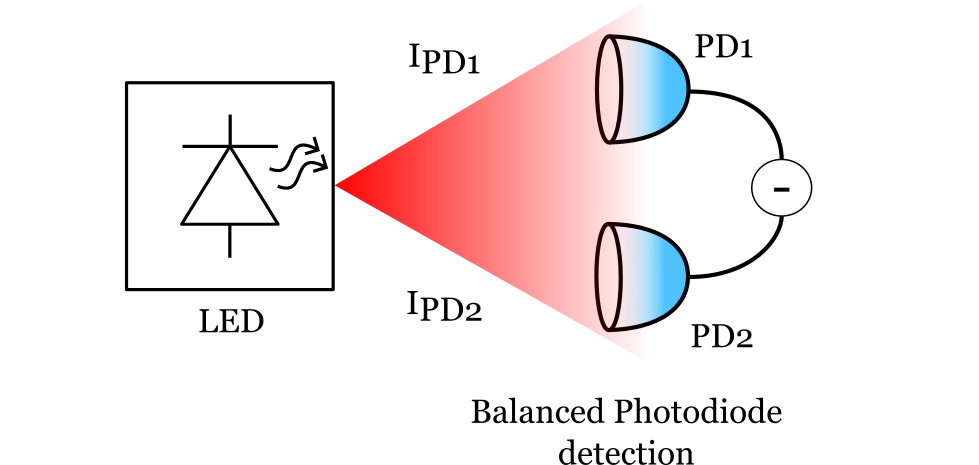}
    \caption{The balanced detection scheme for an LED and two spatially separated photodiodes, such that the Intensity of beam $I_{PD1}$ and $I_{PD2}$ is the same. The two photodiodes are electrically connected in a differencing scheme. This optical circuit is encapsulated within our linear optocoupler device.}
    \label{fig:balanced-det-spatial}
\end{figure}

As shown in Figure~\ref{fig:balanced-det-spatial} our light source (an LED) and photodiodes are placed in a configuration such that the intensity falling on detectors PD1 and PD2 is approximately the same. The spatial separation divides the intensity of the optical field from the LED equally between the two photodiodes, with some fraction of photons lost. PD1 and PD2 are connected in a differencing scheme. We can consider the light field from the LED to consist of classical noise attributed to the driving current, and quantum noise attributed to the discretisation of the light field (shot noise). Since the classical noise is not related to photon fluctuations it will be correlated at PD1 and PD2, and quantum noise by definition will be uncorrelated.

It is apparent that the mean photocurrent generated by PD1 and PD2 (denoted $<i_1>$ and $<i_2>$) will be identical, therefore $<i_1>-<i_2>=\epsilon$, where $\epsilon$ is small compared to the measured values and depends on the matching of the photodiodes, and the spatial beam shape of the LED. Classical fluctuations are therefore suppressed from the output signal. Since the shot noise component of the two beams is independent, the photocurrent fluctuations in $i_1$ and $i_2$ will be completely uncorrelated and the noise power will add up at the output (voltages will sum as root square $\sqrt{V_1^2+V_2^2...}$). The combined shot noise in $i_1$ and $i_2$ must have the same magnitude as that from a single photodiode detecting the whole intensity from the LED. Therefore the output of the balanced photodiode is the same as the shot noise of the LED detected by a single photodiode~\cite{Balanced-det-ch-fox}.

For photodiode of quantum efficiency $\eta$ and photon flux $\Phi$, the photocurrent is
\begin{equation}
    i=\eta e\Phi=\eta e\frac{P}{\hbar\omega}
\end{equation}
\noindent where e is the charge of an electron, P is the power of the light beam, and $\omega$ is the angular frequency.

The photocurrent produced by the photodiode will fluctuate because of the underlying fluctuations in the impinging photon number. Photon number fluctuations are reflected in photocurrent fluctuations with a fidelity determined by $\eta$. The time-varying photocurrent i(t) can be broken into a time-independent average current $<i>$ and time-varying fluctuations $\Delta i(t)$
\begin{equation}
    i(t)=<i>+\Delta i(t)
\end{equation}

The average value of $\Delta i(t)$ is zero but the average of square of $\Delta i$, $<(\Delta i(t)^2)>$, will not be zero. If we consider a load resistor $R_L$ for the photodiode measuring shot noise, the time-varying noise power becomes,
\begin{equation}
    P_{noise}(t)=(\Delta i(t))^2R_L
\end{equation}
for an optical field with Poisson photon number statistics.
\begin{equation}
    P(n) = \frac{\bar n^n}{n!}e^{-\bar n},n=0,1,2.......
\end{equation}
    and
\begin{equation}
    (\Delta n)^2=\bar n
\end{equation}
where $\Delta n$ is standard deviation in the number of photons and $\bar n$ is the mean.
The photoelectron statistics would also be expected to follow a Poisson distribution,
\begin{equation}
    (\Delta N)^2 = <N>
\end{equation}

Since i(t) is proportional to the number of photoelectrons generated per second, the photocurrent variance will satisfy,
\begin{equation}
    (\Delta i)^2 \propto <i>
\end{equation}

It should be clear from the above relation that the variance of current fluctuations in the photocurrent is directly proportional to the average value of the photocurrent \cite{shot-noise-photodiode-ch-fox}. Since we are using a balanced scheme, the variance of the photon number distribution impinging on the two photodiodes sums, while the mean is reduced to zero. Therefore at the output, we will have a distribution such that the variance would increase linearly with a linear increase in the intensity of the optical field.

\section{Experiment and Results}

\begin{figure*}
    \centering
    \includegraphics[width=\linewidth]{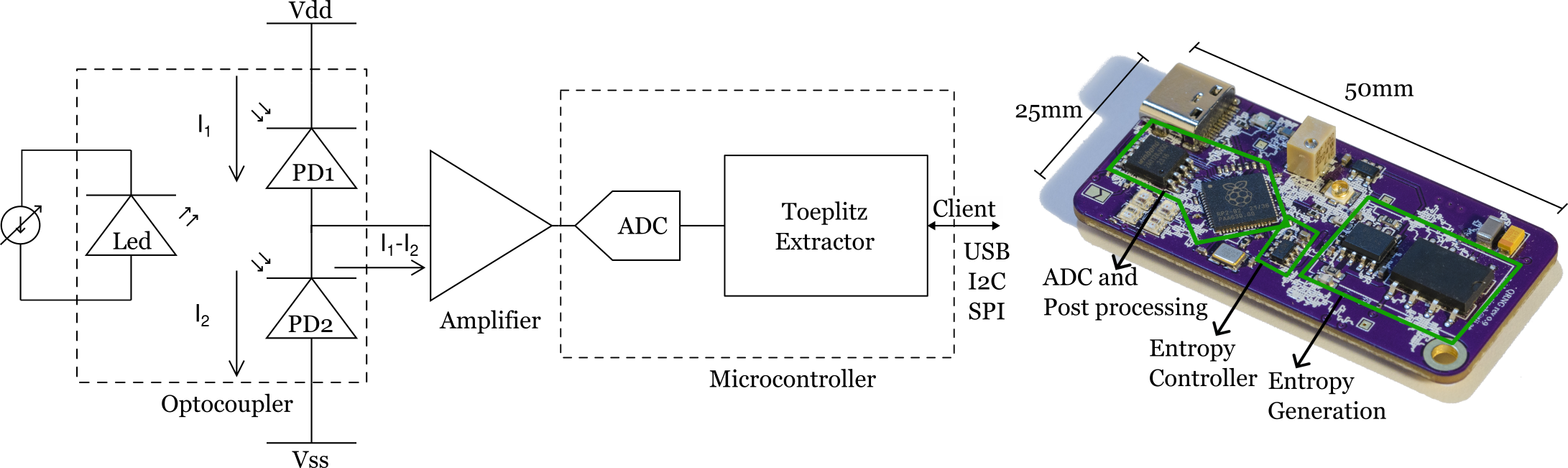}
    \caption{Experimental setup, showing the photodiode connection, the amplification scheme and the ADC and extractor for conditioning of the raw entropy stream (block diagram). We also show a photograph of our device, with call-outs for the primary blocks. This device was used to produce all random data analysed in this manuscript.}
    \label{fig:balanced-detection}
\end{figure*}

\noindent We design a printed circuit board with all necessary components for the QRNG in a 25x50mm form factor as shown in figure~\ref{fig:balanced-detection}. We use this evaluation board as our experimental setup for this section. The photodiodes PD1 and PD2 generate a shot noise signal at the output. The bandwidth of the integrated photodiodes in the linear optocoupler we use is 250 KHz. Shot noise manifests as a very weak current noise signal and therefore needs high amplification and conversion to voltage for further processing. We divide the amplification into two stages to preserve the bandwidth while providing a high gain to utilise the ADC properly. We combine a transimpedance amplifier in the first stage with a very high transimpedance gain and a voltage amplifier to further amplify the signal to be captured with an ADC. A fixed offset of $V_{dd}/2$ is provided to the noise signal for optimal digitisation.
We use the integrated ADC from the microcontroller RP2040 to capture the signal. RP2040 has a 12-bit successive approximation ADC with a Maximum sampling rate of 500 KSa/s, an Effective Number of Bits (ENOB) of 8 bits and a SNR of 53 dB~\cite{Omo_2021,rp2040_datasheet}. Since the sampling rate of the ADC is much higher than the bandwidth of the photodiodes we configure the ADC to run at a sub-nyquist sampling rate (100 KSa/s) to ensure independence of the sampled events (IID criterion). It is important to recall that sampling at a rate higher or equal to the bandwidth of the signal will lead to correlation in successive output bits and therefore running the ADC at a lower sampling rate is required to eliminate these correlations. The ADC then converts the analog signal to digital codes that can then be run through an extractor algorithm. To collect data for this manuscript, We transfer the raw ADC codes to a computer on which we implement a Toeplitz extractor. Note that we can run an instance of Toeplitz extractor on the microcontroller directly, enabling standalone operation. In this case however, the modest processing power of the RP2040 leads to low output rate which may nevertheless be considered sufficient for some applications.

We classify the noise signal at the output of the voltage amplifier into two categories, quantum noise and classical noise. Quantum noise is defined here as the noise generated by the optocoupler which corresponds to the shot noise of the LED. Classical noise is any other noise contribution, this includes noise such as thermal noise, detector dark noise, and amplifier noise. For shot noise, we characterise the output from our setup to confirm two points:
\begin{enumerate}
    \item The contribution of shot noise to the output signal at the voltage amplifier is much more significant than classical noise from the system. We quantify this by calculating the Quantum to classical noise ratio (QCNR), i.e. the ratio of shot noise from the optocoupler to the overall classical noise in the system. We also calculate the min-entropy which is the minimum entropy of the raw signal.
    \item The output distribution from the voltage amplifier is Poissonion, important to prove that the distribution corresponds to the shot noise and that this noise dominates the output statistics.
\end{enumerate}

\begin{figure}
    \centering
    \includegraphics[width=\linewidth]{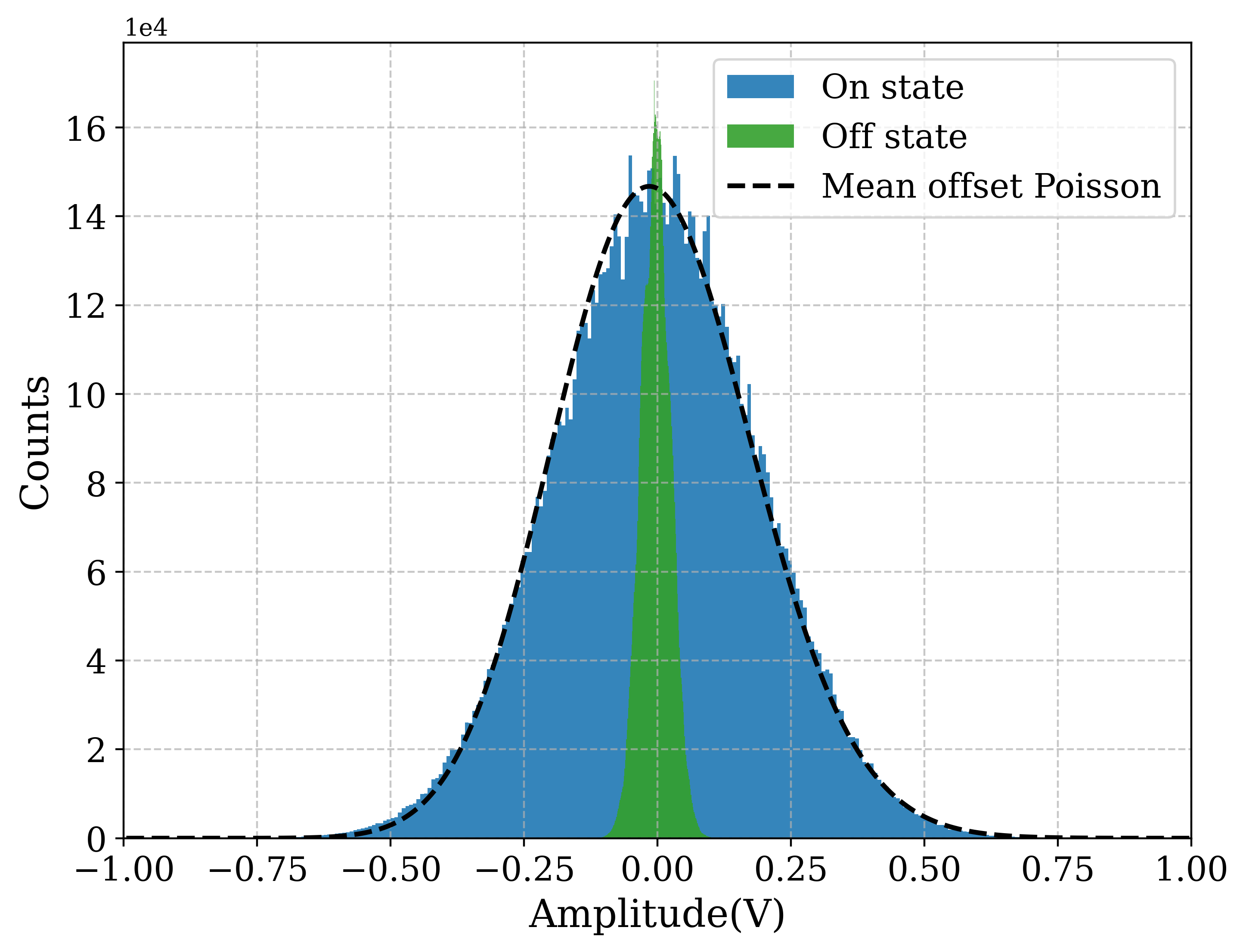}
    \caption{Histogram of the QRNG observed with an oscilloscope with the current source for the LED off(Green) and with the current source providing 70mA of forward current to the LED(Blue). We observe that the variance of the noise when the LED is on is significantly larger than with the LED off.}
    \label{fig:poisson-dist}
\end{figure}

In Figure~\ref{fig:poisson-dist}, we show the output statistics of the noise signal captured by the oscilloscope. We capture the statistics with the LED off(green), this is the overall classical noise in the system. We then turn on the LED of the optocoupler and capture the statistics(blue), this is the mixture of quantum and classical noise in the system. Since the variance of the quantum noise is much more significant than the classical noise, we can say that the quantum noise in the output signal dominates. To calculate the QCNR, we use the variance of Classical noise ($\sigma_{C}^2$) and quantum noise($\sigma_{Q}^2$) and use the equation,
\begin{equation}
    QCNR=20log(\frac{\sigma_{Q,C}^2-\sigma_{C}^2}{\sigma_{C}^2})
\end{equation}
where $\sigma_{Q,C}^2$ is the variance of the output noise signal with the LED ON given in blue in Figure~\ref{fig:poisson-dist} and $\sigma_{C}^2$ is the variance of the output noise signal with the LED OFF given in green. The QCNR for our device is 32\,dB. Quantum noise in the system dominates the overall signal at the output.

It is also important to calculate the min-entropy of the output sequence. The min-entropy is the minimum information-theoretic random bits in a sequence. The minimum entropy is used to define the extraction ratio which bounds the maximum output bits from the extractor. We use a Toeplitz extractor to extract information theoretic random numbers from the output digital codes. To calculate the min-entropy from the output statistics, we use the equation,
\begin{equation}
    H_{\infty}(X)=-log_2(\underset{x\in \{0,1\}^n}{max}P_r[X=x])
\end{equation}

It quantifies the amount of randomness of a distribution X on $\{0,1\}^n$~\cite{max_randomness_extractor,min-max-entropy-meaning,fuzzy-extractor}. The entropy of given sequence X is determined by sample point $x$ with maximal probability. We calculate the extraction ratio of our device to be 7 bits for the 12-bit ADC on RP2040. Putting a sufficient clearance to this limit, we use a Toeplitz extractor to extract 5 bits per sample.
\begin{figure}
    \centering
    \includegraphics[width=\linewidth]{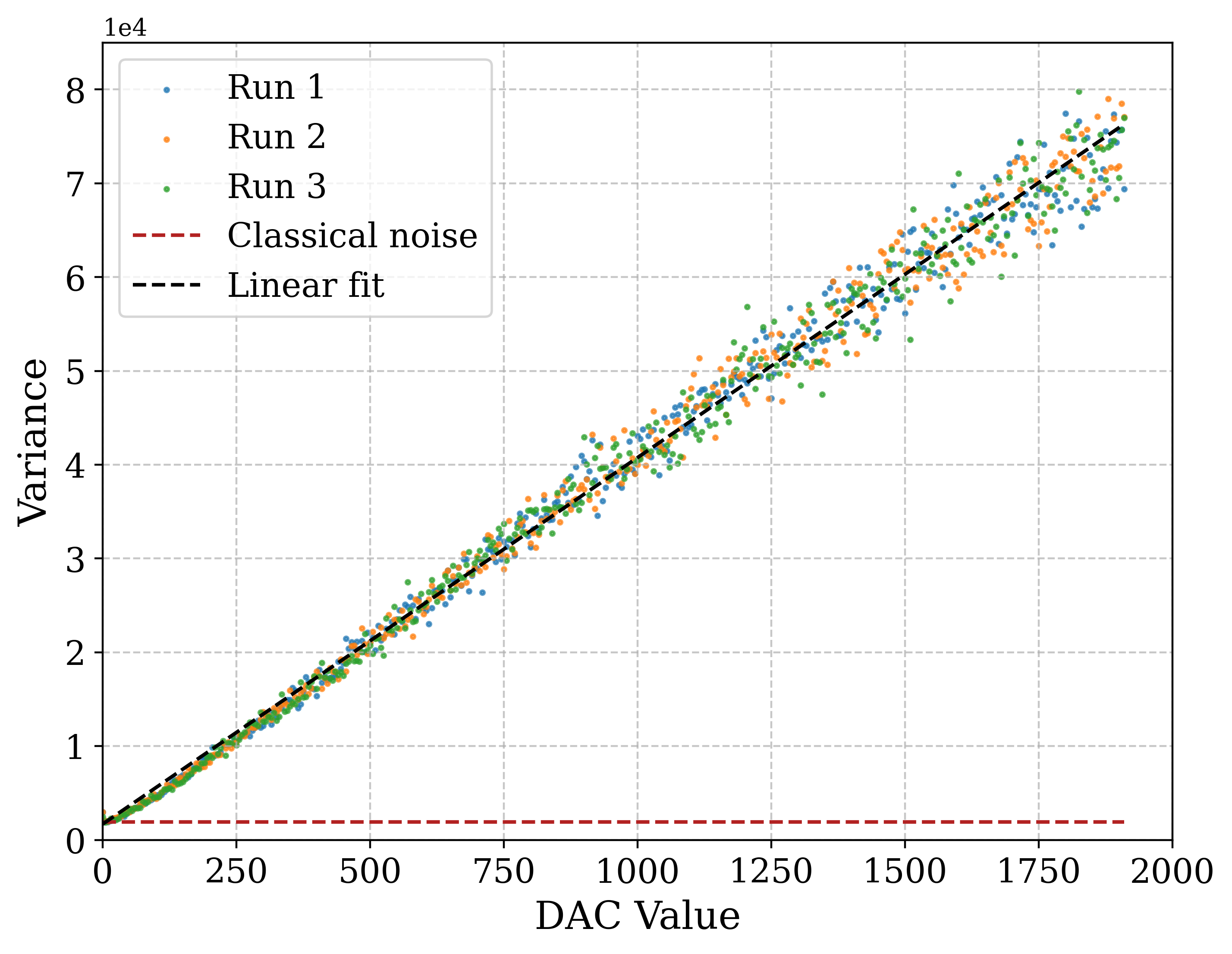}
    \caption{Variance of the output noise signal for different values of forward current to the LED. We evaluate three cycles, varying the value of the DAC-controlled current source from 0 to 2000 to check that the Poisson characteristic does not change. The data supports a linear fit (trend line), from which we confirm the variance change is linear with the change of the mean value of the optical field.}
    \label{fig:linear-3run}
\end{figure}

In Figure~\ref{fig:linear-3run}, we vary the forward current of the LED and measure the variance of the output signal. Changing the forward current of the LED linearly increases the intensity of the optical field, since photodiodes are linear, this manifests as a linear change in the mean of the output signal. Since we are using a balanced scheme this mean is cancelled out electrically and we only have access to the noise signal. For a Poisson distribution, the variance of the output noise signal should vary linearly with the mean and therefore the forward current. We have an onboard DAC-controlled current source that can be controlled using the microcontroller. By linearly stepping the DAC value, we linearly increase the forward current. We can see in Figure~\ref{fig:linear-3run} that the variance value is increasing with a linear trend for three separately acquired datasets. This trend would be quadratic for super Poissonian distribution if the classical noise in the system dominates.
\begin{figure}
    \centering
    \includegraphics[width=\linewidth]{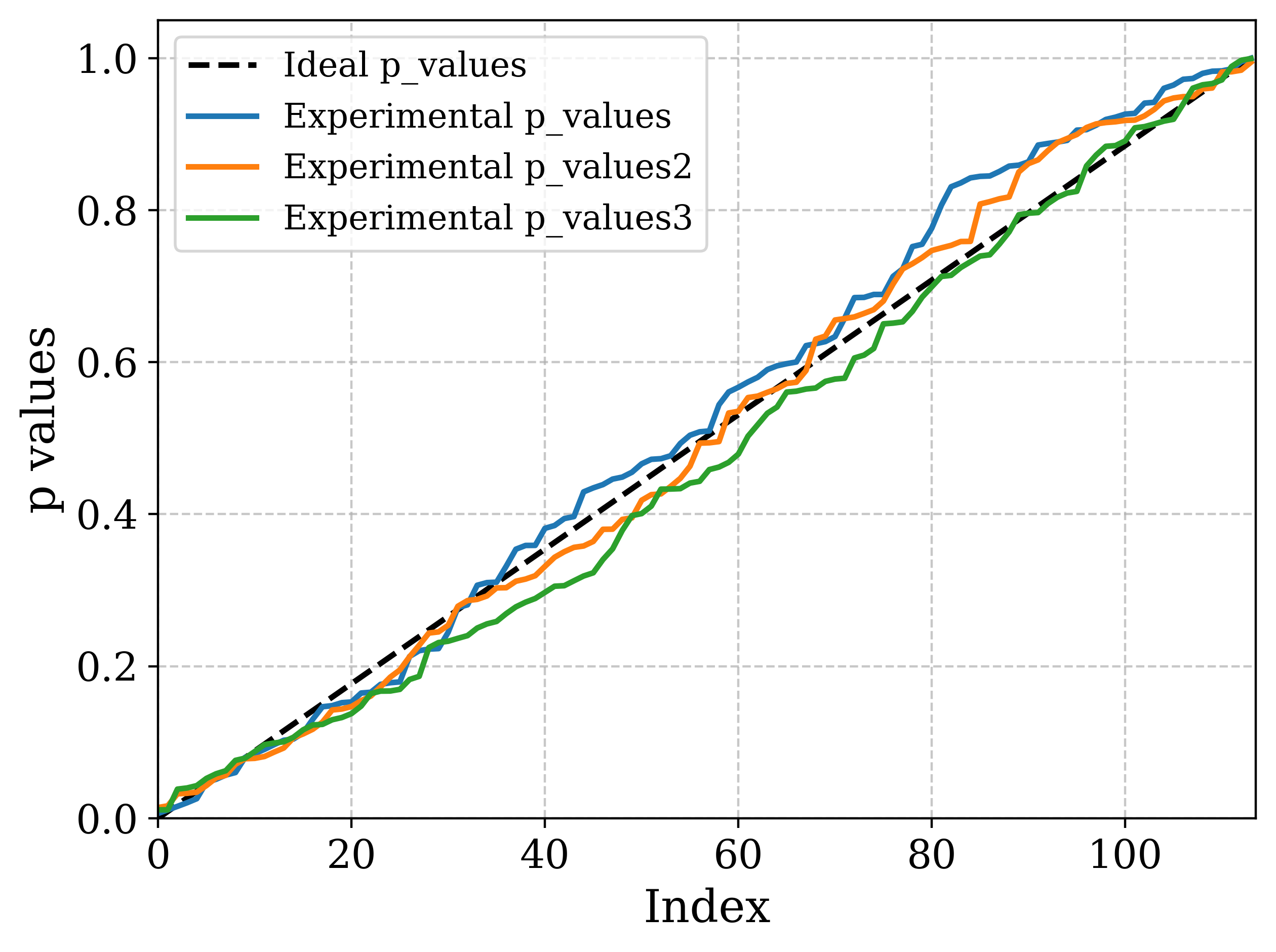}
    \caption{KS plot of the p-values from the output of a dieharder test run on 900\,Mbit data, divided into three chunks and tested individually for statistical randomness. The output of the KS test should be a uniform distribution of p-values for truly random bit, shown by the dotted line labelled ideal p-values.}
    \label{fig:dieharder-test-ks}
\end{figure}

The data presented in Figure~\ref{fig:poisson-dist} is used to determine the minimum entropy in the output sequence, and the overall QCNR. Datasets such as those plotted in Figure~\ref{fig:linear-3run} are further used to confirm that the distribution is in agreement with a Poissonian model. We collect 1.2\,Gb of random bits and divide it into four batches of 300\,Mb each at an acquisition rate of 181\,Kb/sec of raw entropy and then run a Toeplitz extractor on the captured data on the computer. We note that the rate of the QRNG is limited by the choice of ADC and microcontroller. The theoretical limit for random number generation can be calculated for this design as 1.2\,Mb/s, sampling at 250\,KSa/s.

\begin{figure}
    \centering
    \includegraphics[width=\linewidth]{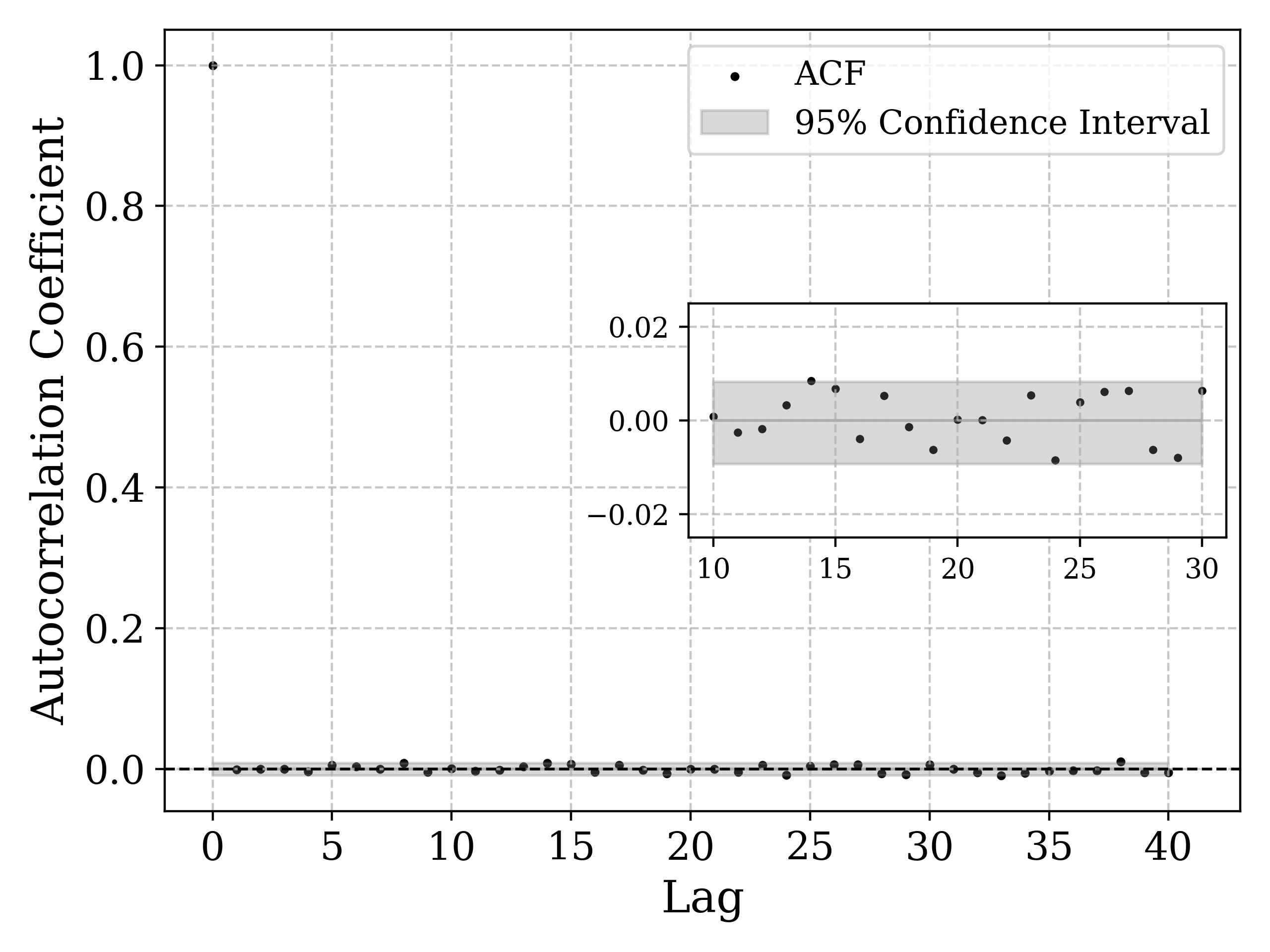}
    \caption{Autocorrelation function run with bitwise lag for a sequence of 5\,Mbits. The peak at 5\,Mbits shows the autocorrelation of the sequence with itself at zero lag.}
    \label{fig:autocorrelation-test}
\end{figure}

Testing randomness is known to be non-trivial, and most tests rely on running some variation of statistical analysis on the output of the random generator. Passing these testing suites is necessary but insufficient to evaluate the quality of your randomness. To test the statistical randomness of the generated output we run our extracted data through the ``dieharder'' test suite~\cite{dieharder} and autocorrelation tests. The result of the dieharder test is visualised in the KS test shown in Figure~\ref{fig:dieharder-test-ks}~\cite{K-S_test_dieharder,dieharder-github}. In this analysis, we compare the Cumulative Density Function (CDF) of the p-values from our test to the CDF of an ideal distribution of p-values for statistically random data.  We show that the trend for each run of the 300\,Mb file is linear, as expected for a truly uniform output.
We then compute the autocorrelation of 5\,Mbit of extracted data. This is calculated by performing a cross-correlation of a sequence with itself, lagged bitwise. If x and y are 1-D arrays, the equation used for the cross-correlation is,
\begin{equation}
    Z[k]=(x*y)(k-N+1)=\sum_{l=0}^{||x||-1}x_l y_{l-k+N-1}
\end{equation}

for $k=0,1.......,||x||+||y||-2$, where $||x||$ is the length of x, $N=max(||x||,||y||)$, and $y_{m}$ is 0 when m is outside the range of y~\cite{2020SciPy-NMeth}. The result is plotted in Figure~\ref{fig:autocorrelation-test}.

Any correlation in the sequence would show as peaks in Figure~\ref{fig:autocorrelation-test}. We observe only one peak corresponding to zero lag, or alternatively the cross-correlation of the sequence with itself. Correlations within data chunks would lead to points lying outside the three-sigma bound for the correlation coefficients. As we increase the size of the data to run the correlation test the value of the correlation coefficients would also decrease.

\section{Conclusions}

\noindent We describe a scheme to generate quantum random numbers from shot noise using balanced photodetection in commercial off-the-shelf available components. We describe a "white-box" implementation of the QRNG, a practical alternative to the device-independent QRNG approach, where instead of extracting random numbers from a black-box implementation of a trusted QRNG manufacturer or very complicated device-independent QRNG, the OEM manufacturer can include a reference QRNG design into their equipment with commercially available components.  Following this logic, the scheme has been optimised for low cost, small footprint\,(25x50mm), and ease of manufacture. We develop a prototype QRNG device and describe experiments that can be used to validate its operation. We further implement several methods to characterise the noise of the system, without relying on direct access to the optical field. We calculate the QCNR to be above 30\,dB. The raw entropy rate is 181\,Kb/s. The rate of the QRNG is sufficient for post-quantum cryptography implementation on personal devices such as laptops.

Adopting a higher bandwidth optocoupler, a faster ADC and microcontroller we expect to increase the output rate further. By way of illustrative example, for a 1\,MHz bandwidth linear optocoupler, a 750\,KSa/s ADC and a 7-bit extraction ratio, we would expect to achieve 5.25\, Mbit/s of entropy. All of these components are easily available from different manufacturers within the commercial off-the-shelf electronics supply chain. We believe that QRNG devices of the kind described here will be an attractive option for integration into many electronics-based platforms that currently lack a robust and reliable entropy source.

\newpage
\bibliographystyle{ieeetr} 
\bibliography{ref.bib}

\begin{thebibliography}{10}

\bibitem{randomness_crypto}
R.~Gennaro, ``Randomness in cryptography,'' {\em IEEE Security \& Privacy}, vol.~4, no.~2, pp.~64--67, 2006.

\bibitem{quantum_lottery}
S.~Mishra and A.~Pathak, ``Quantum and semi-quantum lottery: strategies and advantages,'' {\em Quantum Information Processing}, vol.~22, p.~290, July 2023.

\bibitem{CHSH_randomness}
X.~Yuan, Z.~Cao, and X.~Ma, ``Randomness requirement on the clauser-horne-shimony-holt bell test in the multiple-run scenario,'' {\em Phys. Rev. A}, vol.~91, p.~032111, Mar 2015.

\bibitem{monte_carlo_method}
N.~Metropolis and S.~Ulam, ``The monte carlo method,'' {\em Journal of the American Statistical Association}, vol.~44, no.~247, pp.~335--341, 1949.
\newblock PMID: 18139350.

\bibitem{random_algorithms}
R.~M. Karp, ``An introduction to randomized algorithms,'' {\em Discrete Applied Mathematics}, vol.~34, no.~1, pp.~165--201, 1991.

\bibitem{quantumtech_telecomm}
V.~Martin, J.~Brito, C.~Escribano, M.~Menchetti, C.~White, A.~Lord, F.~Wissel, M.~Gunkel, P.~Gavignet, N.~Genay, and et~al., ``Quantum technologies in the telecommunications industry,'' {\em EPJ Quantum Technology}, vol.~8, no.~1, 2021.

\bibitem{qrng_cryptography}
M.~Stipčević, ``Quantum random number generators and their use in cryptography,'' in {\em 2011 Proceedings of the 34th International Convention MIPRO}, pp.~1474--1479, 2011.

\bibitem{qrng_review}
M.~Herrero-Collantes and J.~C. Garcia-Escartin, ``Quantum random number generators,'' {\em Reviews of Modern Physics}, vol.~89, Feb. 2017.

\bibitem{carlosqrngquside}
M.~W. Mitchell, C.~Abellan, and W.~Amaya, ``Strong experimental guarantees in ultrafast quantum random number generation,'' {\em Phys. Rev. A}, vol.~91, p.~012314, Jan 2015.

\bibitem{yury_qrng_phase_diff}
R.~Shakhovoy, D.~Sych, V.~Sharoglazova, A.~Udaltsov, A.~Fedorov, and Y.~Kurochkin, ``Quantum noise extraction from the interference of laser pulses in an optical quantum random number generator,'' {\em Opt. Express}, vol.~28, pp.~6209--6224, Mar 2020.

\bibitem{qrng_phase_fluc}
F.~Xu, B.~Qi, X.~Ma, H.~Xu, H.~Zheng, and H.-K. Lo, ``Ultrafast quantum random number generation based on quantum phase fluctuations,'' {\em Optics Express}, vol.~20, no.~11, p.~12366, 2012.

\bibitem{christianqrngvacuum}
Y.~Shi, B.~Chng, and C.~Kurtsiefer, ``{Random numbers from vacuum fluctuations},'' {\em Applied Physics Letters}, vol.~109, p.~041101, 07 2016.

\bibitem{100gbpsvacuumqrng}
C.~Bruynsteen, T.~Gehring, C.~Lupo, J.~Bauwelinck, and X.~Yin, ``100-gbit/s integrated quantum random number generator based on vacuum fluctuations,'' {\em PRX Quantum}, vol.~4, p.~010330, Mar 2023.

\bibitem{qrng_vacuum_fluc}
C.~Bruynsteen, T.~Gehring, C.~Lupo, J.~Bauwelinck, and X.~Yin, ``100-gbit/s integrated quantum random number generator based on vacuum fluctuations,'' {\em PRX Quantum}, vol.~4, no.~1, 2023.

\bibitem{qrngonmobileidq}
B.~Sanguinetti, A.~Martin, H.~Zbinden, and N.~Gisin, ``Quantum random number generation on a mobile phone,'' {\em Phys. Rev. X}, vol.~4, p.~031056, Sep 2014.

\bibitem{ID_Quantique_2022_QRNG}
{ID Quantique SA}, ``Only quantum random number generators are intrinsically random and provably unpredictable,'' Jun 2022.
\newblock Available at \url{https://www.idquantique.com/random-number-generation/products/}.

\bibitem{Quside_2024}
{Quside Technologies}, ``{QRNG Security Chipset - QN100 Chip}.'' \url{https://quside.com/product/quside-qn100-chipsets/}, Feb 2024.

\bibitem{laser_noise_supression}
G.~Abbas, V.~Chan, and T.~Yee, ``A dual-detector optical heterodyne receiver for local oscillator noise suppression,'' {\em Journal of Lightwave Technology}, vol.~3, no.~5, pp.~1110--1122, 1985.

\bibitem{balanced_photodetector}
Z.~Li, H.~Chen, H.~Pan, A.~Beling, and J.~C. Campbell, ``High-power integrated balanced photodetector,'' {\em IEEE Photonics Technology Letters}, vol.~21, no.~24, pp.~1858--1860, 2009.

\bibitem{led_qrng_entropy}
G.~Gras, A.~Martin, J.~W. Choi, and F.~Bussières, ``Quantum entropy model of an integrated quantum-random-number-generator chip,'' {\em Physical Review Applied}, vol.~15, no.~5, 2021.

\bibitem{perovskite_led_qrng}
J.~Argillander, A.~Alarcón, C.~Bao, C.~Kuang, G.~Lima, F.~Gao, and G.~B. Xavier, ``Quantum random number generation based on a perovskite light emitting diode,'' {\em Communications Physics}, vol.~6, no.~1, 2023.

\bibitem{Optocoupler_qrng}
Y.-Y. Hu, Y.-Y. Ding, S.~Wang, Z.-Q. Yin, W.~Chen, D.-Y. He, W.~Huang, B.-J. Xu, G.-C. Guo, and Z.-F. Han, ``Compact quantum random number generation using a linear optocoupler,'' {\em Optics Letters}, vol.~46, no.~13, p.~3175, 2021.

\bibitem{sub-shot-noise-imaging}
G.~Brida, M.~Genovese, and I.~Ruo~Berchera, ``Experimental realization of sub-shot-noise quantum imaging,'' {\em Nature Photonics}, vol.~4, p.~227–230, Feb. 2010.

\bibitem{shot-noise-detector}
G.~Brida, M.~Genovese, and I.~Ruo~Berchera, ``Experimental realization of sub-shot-noise quantum imaging,'' {\em Nature Photonics}, vol.~4, pp.~227--230, Apr. 2010.

\bibitem{shot-noise-photodiode-ch-fox}
M.~Fox, {\em Shot noise in Photodiodes}, p.~94–99.
\newblock Oxford University Press, 2013.

\bibitem{squeezed-light}
D.~F. Walls, ``Squeezed states of light,'' {\em Nature}, vol.~306, pp.~141--146, Nov. 1983.

\bibitem{Balanced-det-ch-fox}
M.~Fox, {\em Detection of squeezed light}, p.~139–142.
\newblock Oxford University Press, 2006.

\bibitem{Omo_2021}
M.~Omo, ``Introduction,'' Feb 2021.
\newblock Available at \url{https://pico-adc.markomo.me/}.

\bibitem{rp2040_datasheet}
{Raspberry Pi}, ``{RP2040 Datasheet A microcontroller by Raspberry Pi}.''
\newblock Available at \url{https://datasheets.raspberrypi.com/rp2040/rp2040-datasheet.pdf}.

\bibitem{max_randomness_extractor}
J.~Y. Haw, S.~M. Assad, A.~M. Lance, N.~H.~Y. Ng, V.~Sharma, P.~K. Lam, and T.~Symul, ``Maximization of extractable randomness in a quantum random-number generator,'' {\em Phys. Rev. Appl.}, vol.~3, p.~054004, May 2015.

\bibitem{min-max-entropy-meaning}
R.~Konig, R.~Renner, and C.~Schaffner, ``The operational meaning of min- and max-entropy,'' {\em IEEE Transactions on Information Theory}, vol.~55, no.~9, pp.~4337--4347, 2009.

\bibitem{fuzzy-extractor}
Y.~Dodis, R.~Ostrovsky, L.~Reyzin, and A.~Smith, ``Fuzzy extractors: How to generate strong keys from biometrics and other noisy data,'' {\em SIAM Journal on Computing}, vol.~38, no.~1, pp.~97--139, 2008.

\bibitem{dieharder}
R.~G. Brown, ``Dieharder: A random number test suite.''
\newblock Available at \url{https://webhome.phy.duke.edu/~rgb/General/dieharder.php}.

\bibitem{K-S_test_dieharder}
F.~J.~M. Jr., ``The kolmogorov-smirnov test for goodness of fit,'' {\em Journal of the American Statistical Association}, vol.~46, no.~253, pp.~68--78, 1951.

\bibitem{dieharder-github}
r.~Jochen~Voss, ``dieharder.''
\newblock Available at \url{https://github.com/seehuhn/dieharder}.

\bibitem{2020SciPy-NMeth}
P.~Virtanen, R.~Gommers, T.~E. Oliphant, M.~Haberland, T.~Reddy, D.~Cournapeau, E.~Burovski, P.~Peterson, W.~Weckesser, J.~Bright, S.~J. {van der Walt}, M.~Brett, J.~Wilson, K.~J. Millman, N.~Mayorov, A.~R.~J. Nelson, E.~Jones, R.~Kern, E.~Larson, C.~J. Carey, {\.I}.~Polat, Y.~Feng, E.~W. Moore, J.~{VanderPlas}, D.~Laxalde, J.~Perktold, R.~Cimrman, I.~Henriksen, E.~A. Quintero, C.~R. Harris, A.~M. Archibald, A.~H. Ribeiro, F.~Pedregosa, P.~{van Mulbregt}, and {SciPy 1.0 Contributors}, ``{{SciPy} 1.0: Fundamental Algorithms for Scientific Computing in Python},'' {\em Nature Methods}, vol.~17, pp.~261--272, 2020.

\end{thebibliography}

\end{document}